\documentclass[9pt,conference]{IEEEtran}
\IEEEoverridecommandlockouts
\usepackage{cite}
\usepackage{amsmath,amssymb,amsfonts}
\usepackage{algorithmic}
\usepackage{graphicx}
\usepackage{textcomp}
\usepackage{xcolor}
\usepackage{setspace}
\usepackage[small]{caption}
\usepackage{fancyvrb}
\usepackage{pgfplots}
\usepackage{multirow}
\usepackage{xspace}

\usepgfplotslibrary{external}
\pgfplotsset{compat=newest}
\usepgfplotslibrary{statistics}
\usepgfplotslibrary{groupplots}
\usetikzlibrary{patterns}

\definecolor{cb1}{HTML}{7E92B8}
\definecolor{cb2}{HTML}{EDFAC4}
\definecolor{cb3}{HTML}{E0B0D5}
\definecolor{cb4}{HTML}{FFECC8}

\newcommand{\linebreakand}{%
  \end{@IEEEauthorhalign}
  \hfill\mbox{}\par
  \mbox{}\hfill\begin{@IEEEauthorhalign}
}

\newcommand{\aker}{\textsc{Aker}\xspace}

\def\BibTeX{{\rm B\kern-.05em{\sc i\kern-.025em b}\kern-.08em
    T\kern-.1667em\lower.7ex\hbox{E}\kern-.125emX}}
\begin{document}

\title{\aker: A Design and Verification Framework for Safe and Secure SoC Access Control} 

\author{\IEEEauthorblockN{Francesco Restuccia\IEEEauthorrefmark{2}\IEEEauthorrefmark{1},
Andres Meza\IEEEauthorrefmark{1}, and
Ryan Kastner\IEEEauthorrefmark{1}}\\
\IEEEauthorblockA{\IEEEauthorrefmark{1} UC San Diego \quad\quad\quad\quad \IEEEauthorrefmark{2} Scuola Superiore Sant'Anna Pisa}
}

\maketitle
\begin{abstract}

Modern systems on a chip (SoCs) utilize heterogeneous architectures where multiple IP cores have concurrent access to on-chip shared resources.
In security-critical applications, IP cores have different privilege levels for accessing shared resources, which must be regulated by an access control system.
\aker is a design and verification framework for SoC access control. 
\aker builds upon the Access Control Wrapper (ACW) -- a high performance and easy-to-integrate hardware module that dynamically manages access to shared resources. To build an SoC access control system, \aker distributes the ACWs throughout the SoC, wrapping controller IP cores, and configuring the ACWs to perform local access control.  
To ensure the access control system is functioning correctly and securely, \aker provides a property-driven security verification using MITRE common weakness enumerations. \aker verifies the SoC access control at the \emph{IP level} to ensure the absence of bugs in the functionalities of the ACW module, at the \emph{firmware level} to confirm the secure operation of the ACW when integrated with a hardware root-of-trust (HRoT), and at the \emph{system level} to evaluate security threats due to the interactions among shared resources.
The performance, resource usage, and security of access control systems implemented through \aker is experimentally evaluated on a Xilinx UltraScale+ programmable SoC, it is integrated with the OpenTitan hardware root-of-trust, and it is used to design an access control system for the OpenPULP multicore architecture.

\end{abstract}

\section{Introduction}

Modern System on a Chip (SoC) have heterogeneous architectures comprised of microprocessors, hardware accelerators, on-chip memory hierarchies, and I/O.  They utilize complex on-chip communication networks where the  processors and  accelerators transfer information between themselves and other shared resources, often with tight constraints on throughput, latency, and resource usage.

In security-critical applications, on-chip resources have different levels of trustworthiness and criticality that are often dynamic in nature. Examples include: 1) a shared memory may be (temporally) isolated from an untrusted IP core, 2) certain resources are only accessible during debug mode, and 3) only the hardware root of trust can access security critical control and status registers. 

In order to operate in a safe and secure manner, SoCs use an \emph{access control system} that enforces an \emph{access control policy}. The access control policy defines the ability of the SoC controllers to access the different peripherals. The access control policy changes over the SoC life-cycle -- design, manufacturing, testing, passing through several OEMs, and on to the final user.  Access control policies are dynamic, e.g., policies differ when in boot mode, secure operating modes, reset, and normal operating scenarios. Thus, SoC access control policies require 
an efficient and flexible access control system.

The access control system plays a critical role for ensuring  safe and secure operation. Thus, it is important that any access control system undergo a rigorous verification process. Verification includes functional correctness. Additionally, and equally as important, it must undergo a security verification process that addresses potential security weaknesses or vulnerabilities. An exploit in the access control system endangers the confidentiality, integrity, and availability of the SoC.

Unfortunately, it is challenging to correctly implement SoC access control systems. The MITRE common weakness enumeration (CWE) database reports a substantial and growing number of hardware weaknesses~\cite{CWE}. Our security verification process identified 30 of these CWEs related to access control systems (see Section~\ref{s:aker}).  
Access control flaws are extremely dangerous as they provide the opportunity for low-level system access. Furthermore, they are challenging to patch. At best they require a firmware rewrite; at worst, they require disabling features or re-manufacturing the chip.

This work proposes \aker~-- a framework for the development of safe and secure on-chip access control systems targeting the requirements of modern safety- and security-critical applications. These requirements include:

\emph{Security Verification: }
\aker provides a property-driven security verification procedure~\cite{hu2016towards} to ensure that the SoC access control policy is devoid of any CWEs. This provides high assurance on the secure operation of \aker-based access control systems. The security verification is done at three levels: the \emph{IP} level, the \emph{firmware} level, and the \emph{system} level. \aker can be easily extended to handle next-generation SoCs and address different CWEs.

\emph{Interoperability:}
\aker is AXI-compliant  and fully transparent to controllers and interconnect. No knowledge or modifications on the internals of the controllers, peripherals, and interconnect are required to integrate an \aker access control system. 
\emph{Immediate Filtering of Illegal Requests:} \aker filters transactions at the source, before entering the interconnect -- no illegal transactions are allowed to enter the network. This avoids any identification issues and prevents system-level interference generated by illegal transactions (e.g., DoS attacks, see Section~\ref{s:fpga-exper}).

\emph{Secure Configuration:}
\aker access control systems easily integrate with a Hardware Root of Trust (HRoT) for runtime monitoring and management of the controllers. Section~\ref{sec:firmware-verification} describes the integration and verification of the OpenTitan~\cite{OpenTitan} HRoT with \aker. 

\emph{Flexibility:}
\aker-based access control systems allows static or dynamic configuration of the access control policy by the HRoT. This provides the required flexibility to cope with the complex life-cycle of modern SoCs.
 
 \emph{Diagnostic Information:}
\aker access control systems log diagnostic information regarding illegal attempts. This provides flexibility to the HRoT on how to perform modules readmission.

\emph{Efficient Performance and Resource Usage:}
\aker access control systems incur only 1 clock cycle delay per AXI transaction independent on the number considered memory regions (see Section~\ref{s:fpga-exper}). This corresponds to an impact of $<1\%$ in all the tested scenarios. 
The resource usage of an \aker-based access control system is minimal and is configurable to be tailored to the SoC and use case.

\emph{Open-Source:}
The design and the security properties developed within \aker are openly released. This allow further verification and provides a solid base for design and security verification extension, facilitating broader use. Our repo contains all the designs, security properties, and security property templates proposed in \aker~\cite{Github}. 

\emph{Ease of Integration:} \aker access control systems are experimentally validated via their integration into  SoCs implemented on a FPGA SoC architecture and on the OpenPULP~\cite{conti2016pulp}. 

\section{SoC Access Control}

An SoC architecture consists of a set of controller devices accessing a set of peripherals devices.\footnote{We adopt the terminology controller/peripheral to describe system-level interactions between IP cores. We use M (manager) and S (subordinate) when specifically referring to the AXI protocol.} 
Different processors, accelerators, and other IP cores can be assigned as a controller. This allows them to autonomously and concurrently communicate with shared peripheral resources available on the SoC, e.g., a DRAM memory controller, on-chip memories, ROM, IP core control and status registers (CSRs), and GPIOs.  
On-chip data transfers use a communication protocols like the AMBA AXI and TileLink, which employ a flexible, asymmetric, synchronous interface targeting high performance and low latency communications. 
A key aspect of any SoC access control system is arbitrating accesses to on-chip resources. High-speed on-chip communications protocols use memory mapped addressing to allow controllers to specify the resources they wish to access. An access control policy specifies whether a data request is allowable at that given time. It is important that the access control system exactly implements the access control policy while having a minimal impact on performance and area.

\subsection{SoC Interconnect Architectures}\label{ss:sample-arch}
\begin{figure}[htb!]
\centering
\includegraphics[width=0.6\columnwidth]{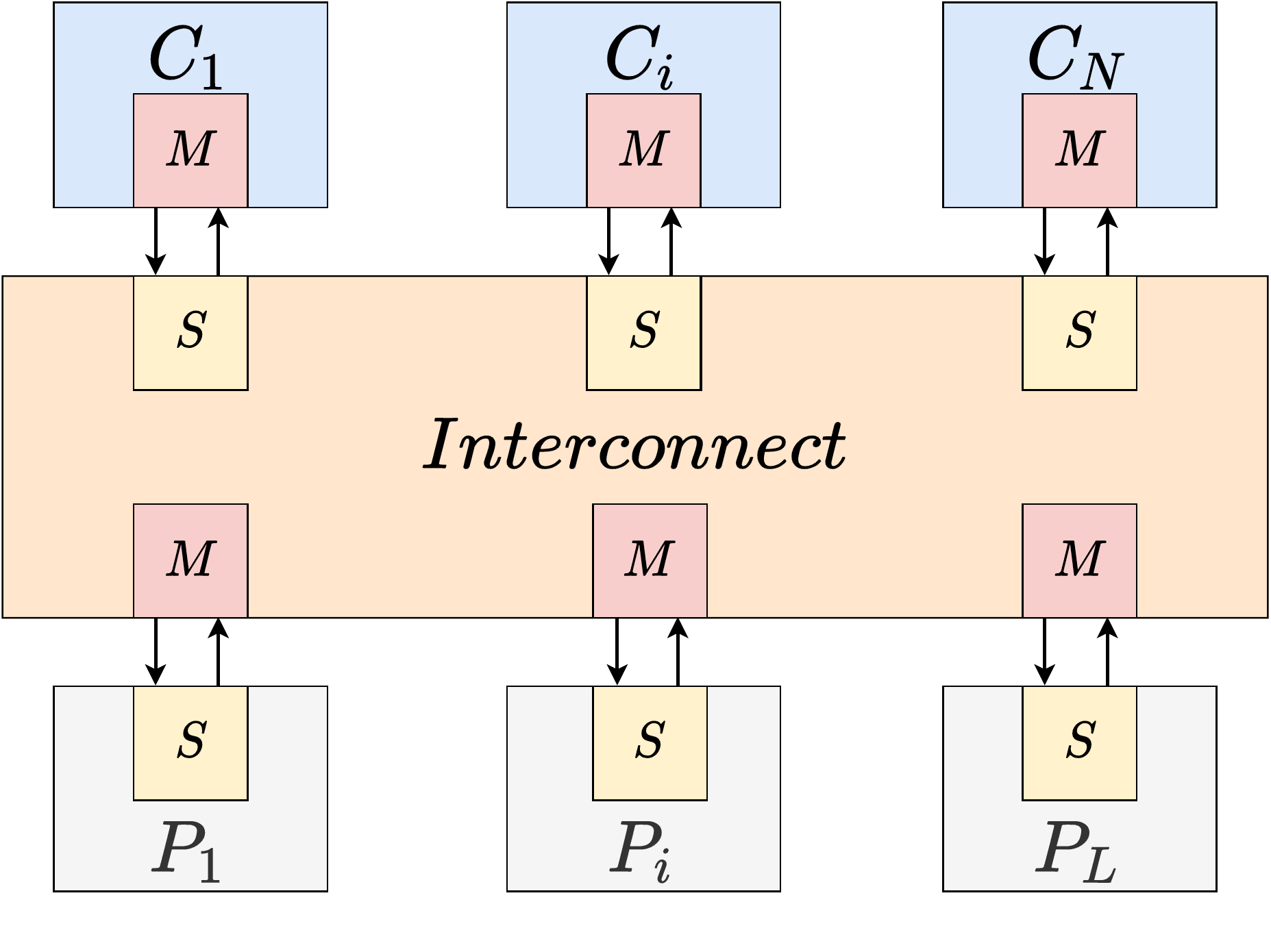}
\caption{An SoC on-chip interconnect architecture composed of $N$ controller modules ($C$) connected to $L$ peripheral modules ($P$).} \label{fig:generic_arch}
\end{figure}

Figure~\ref{fig:generic_arch} shows an SoC interconnect architecture with $N$ controllers $C$ (${C_1, \ldots , C_N}$) each with a manager (M) interface, $L$ peripherals $P$ (${P_1, \ldots , P_L}$) with a subordinate (S) interface, and an interconnect $I_{\text{AXI}}$ connecting them. 
We adopt AMBA AXI standard due to its widespread usage. Our techniques are applicable to other on-chip communication protocols (e.g., TileLink) with minimal modifications. AXI defines an asymmetric communication interface comprised of five independent channels: address read (AR), address write (AW), data read (R), data write (W), and write response (B).
$I_{\text{AXI}}$ arbitrates the access of $C$ modules to the shared $P$ modules.

A controller $C_i$ can initiate a transaction to a shared peripheral $P_j$ issuing an \emph{address request} through its AXI M interface.  %
The address requests is routed to the peripheral $P_j$ by the AXI interconnect.
$P_j$ is accessible by $C_i$ through a unique set of contiguous addresses, also called \emph{peripheral address region}.
$P_j$ serves the received requests providing the required data (read request) or accepting the write data and replying with a write response (write request).

\subsection{Threat Model}\label{ss:threat-model}

The threat model assumes that untrusted controllers $C$ communicate with the on-chip peripherals $P$ using the AXI protocol. 
An access control policy specifies allowable transactions between on-chip controller resources (e.g., microprocessors or hardware accelerators) and shared peripheral resources (e.g., memories and I/O). The policy describes the allowable read and write requests between resources using a set of address ranges. A range is a contiguous address space encoded as the base address and the range size. 
In access control systems implemented in many commercial platforms, security threats are created by the ability of untrusted controllers (e.g., outsourced hardware module IPs) to directly access the peripherals. This is attested by the numerous amount of common weaknesses available in access control systems implemented in industrial applications and enumerated in the MITRE CWE database~\cite{CWE}. 
Typically, such vulnerabilities derive from a superficial \emph{security verification} of the access control system. If properly exploited by controllers, such vulnerabilities enable access to unallowed peripherals, thus violating the access control policy and potentially endangering the confidentiality and integrity of the SoC (see Section~\ref{ss:IP-level-verif}). 
The functionalities of $P$, the routing functionalities of $I_\text{AXI}$, and the Trusted Entity functionalities are assumed to be trusted.

\subsection{On-chip Access Control Systems}
\label{ss:SoA-solutions}

This section discusses options for the implementing on-chip access control system. The primary options include access control monitoring in the interconnect, at the peripherals, in a centralized location, or at the controller (\aker's solution). 

\subsubsection{Access Control using AXI Interconnect}\label{sss:AXI-intc}
The access control policy is enforced within the crossbar interconnect by implementing only the selected physical connections between controllers and peripherals according to the privacy and integrity requirements~\cite{smartconnect}. 
The AXI interconnect is statically configured only with physical connections between a controller and peripheral that are allowed by the policy. 
\textbf{Limitations:}
Hard-coding the access control system does not allow for dynamic updates to the access control policy. 
Therefore, this approach is not a viable option for modern SoCs having complex lifecycles. 
Moreover, the definition of the access control policy can be limited to simplified and unrealistic scenarios for several applications (see Section~\ref{s:fpga-exper}).

\subsubsection{Access Control in Peripherals}\label{sss:peripheral-res}
Each peripheral $P$ includes additional logic that analyzes each request and decides whether to serve it depending on the access control policy~\cite{siddiqui2018pro,coburn2005seca}.
This methodology implicitly assumes that each request is somehow securely marked with the information regarding the identity of the issuing controller. Typically, access control policies enforced in peripherals are configurable, providing some flexibility to handle dynamic policies.

\textbf{Limitations:}  
The AXI standard does not define any information about the source controller.
A common workaround uses the AXI ID signals for identification~\cite{XMPU}. %
However, the AXI IDs are intended to denote parallel execution of threads. An AXI controller is allowed to issue address requests using multiple ID values. Thus, IDs are not suitable for access control.  
Additionally, AXI does not address the integrity of the ID(s), which allows ID manipulation during request propagation and adds uncertainty about the provenance of the request~\cite{amba4spec}.  
Another consideration is that any illegal requests received by a peripheral must be terminated with an AXI-compliant error to avoid network locks. %
This causes unwanted interference with the execution of legal transactions (see Section~\ref{s:fpga-exper}). 
Such solutions can also pose strong limitations in the definition of realistic access control policies (see Section~\ref{s:fpga-exper}).

\subsubsection{Centralized Policy Engine}\label{sss:centralized-policy}

Another option is a centralized security policy engine~\cite{basak2015flexible,nath2018system}. 
The central security policy engine is responsible for authenticating any memory transaction. 

\textbf{Limitations:} All  decisions are made by the central security engine, which requires communication between the wrapper and the security engine on each transaction. This communication  can impact the performance of the system and create bottleneck  at the central security engine. Thus, such solutions are not suitable for high-performance or latency-critical systems.

\section{AKER}\label{s:aker}

\aker is a design and verification framework for developing SoC access control systems, aimed at meeting the performance, security, and flexibility requirements of modern safety- and security-critical applications. 
\aker builds upon the Access Control Wrapper (ACW) -- a high-performance, programmable module that supervises the behaviour of memory transactions from a controller. 
\aker includes property-driven security verification~\cite{hu2016towards} at the IP level, firmware level, and system level. 
\aker uses CWEs to identify potential weaknesses, develop property templates to aid in the property generation process, and use information flow tracking hardware verification tools to validate complex behaviors related to confidentiality and integrity of the SoC access control system.

\subsection{The Access Control Wrapper}\label{ss:intro-acw}

\begin{figure}[htb!]
\centering
\includegraphics[width=0.99\columnwidth]{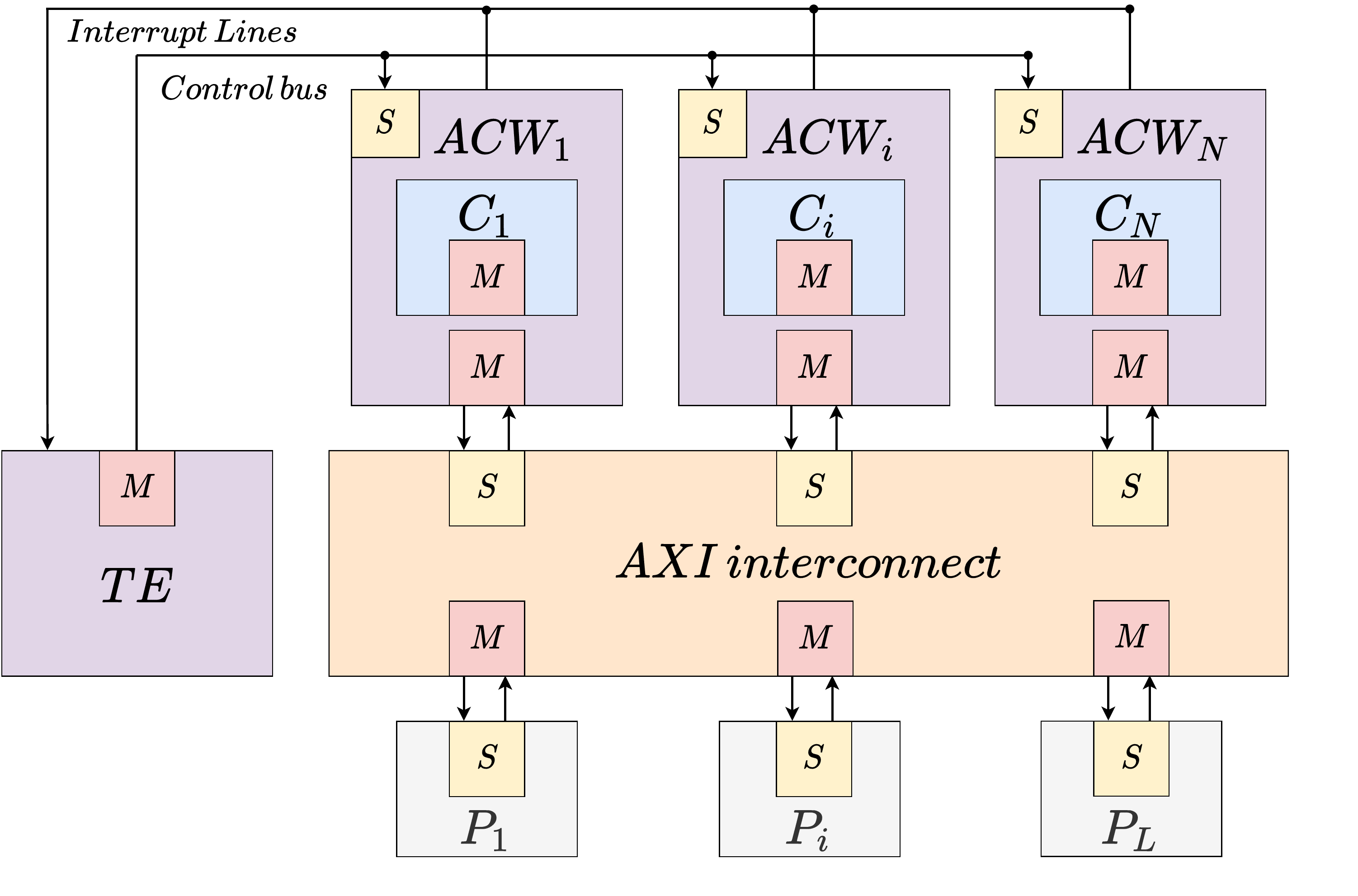}
\caption{Extended AXI multi-controller, multi-peripheral architecture incorporating an \aker-based access control system. The Trusted Entity $TE$ manages the ACW modules. Only legal request are transmitted to the AXI interconnect, i.e., the peripherals receive only legal AXI transactions.} \label{fig:extended_arch}
\end{figure}

The Access Control Wrapper (ACW) is a configurable access control module designed to monitor an AXI-compliant controller.
The ACW exports an AXI M interface, an AXI-lite S configuration interface, and an output interrupt line.
An ACW can be used on any SoC controller resource whose memory accesses require arbitration for safety or security reasons; each  untrusted controller $C_i$ is wrapped by an ACW module $ACW_i$. Figure~\ref{fig:extended_arch} provides an example of an \aker-based access control system.
The M interface of $ACW_i$ is connected to the AXI interconnect (in place of the M interface of $C_i$), while the S interface and the interrupt line are connected to a Trusted Entity ($TE$) (i.e., a HRoT, see Section~\ref{ss:IP-level-verif}). 

$ACW_i$ holds a local access control policy $LACP_i$, configured and maintained by $TE$.
$LACP_i$ describes the address \emph{regions} legally accessible by $C_i$, defining $n_r$ regions for read operations and $n_w$ regions for write operation.  
Each memory request issued by $C_i$ is checked against the configuration of $LACP_i$; if the request is fully contained in at least one of the $LACP_i$'s address regions, the request is considered \emph{legal} by $ACW_i$ and allowed to propagate to the AXI interconnect. 
$n_r$ and $n_w$ impact the resource consumption of the ACW module -- the ACW design allows to easily customize such values according to the SoC requirements. 
To minimize the latency, the address regions are checked in parallel. Thus, the latency introduced by the ACW is constant and independent of $n_r$ and $n_w$. 

The $ACW_i$ has three operating modes:

\emph{1) Reset Mode:} 
the initial state of $ACW_i$. It is awaiting configuration with a valid $LACP_i$. Any request issued by $C_i$ is blocked and does not propagate to the interconnect. Once $LACP_i$ is configured, $ACW_i$ moves to \emph{Supervising Mode}. 

\emph{2) Supervising Mode:} the normal operating mode of the $ACW_i$. Each address request issued by $C_i$ is compared against the stored $LACP_i$. Legal requests are propagated to the AXI interconnect; illegal requests are denied and never reach the AXI interconnect. An illegal request moves $ACW_i$ into \emph{Decouple Mode}.

\emph{3) Decouple Mode:} an illegal request has occurred. $ACW_i$ saves diagnostic information about the illegal request into its internal anomalies registers. $ACW_i$ raises an interrupt to notify the $TE$ of the illegal access attempt. Any further request from $C_i$ is blocked and the $ACW_i$ waits on the $TE$ for readmission.   
Decoupling $C_i$ after an illegal attempt ensures that $TE$ can take appropriate actions on $C_i$ before the safe readmission of the module in the system. 

\emph{Readmission Policy:} The $TE$ can analyze the diagnostic information internal to $ACW_i$ and perform recovery operations on $C_i$ before switching back to \emph{Supervising Mode} and thereby readmitting $C_i$ to communicate to the SoC.
Examples of recovery operations are resetting, reconfiguring, or even reprogramming $C_i$. 
In the most extreme scenario, if the $TE$ decides that the illegal request is the result of a permanent fault, it can keep the $ACW_i$ in \emph{Decouple Mode}, thus permanently disconnecting $C_i$ from the system.

\begin{figure}[htb!]
\centering
\includegraphics[width=0.99\columnwidth]{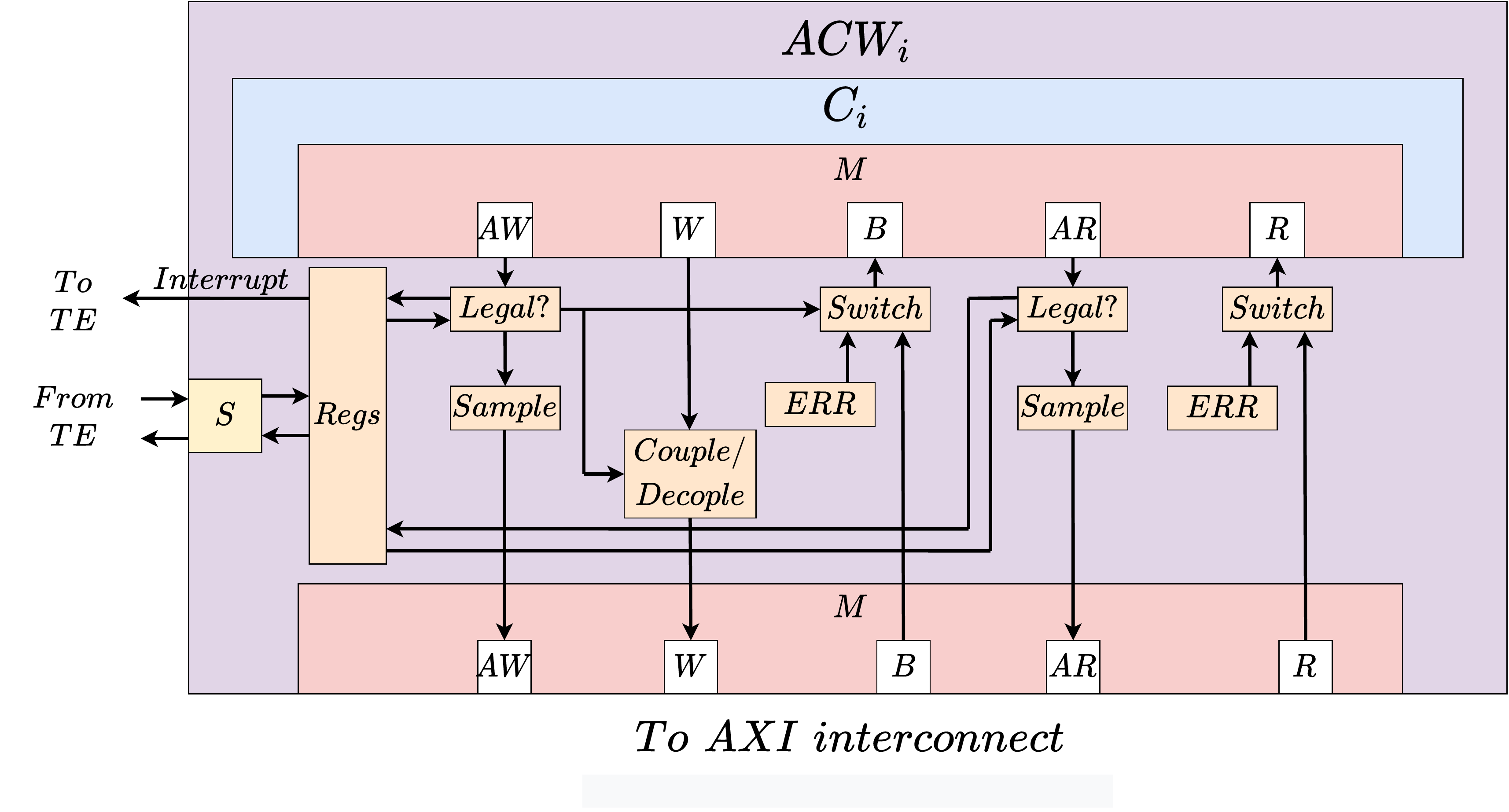}
\caption{$ACW_i$ architecture: $C_i$ is the controller module using an AXI M interface. $Regs$ are the configuration registers holding $LACP_i$. The AXI S interface is connected to the HRoT.}
\label{fig:internals}
\end{figure}

Figure~\ref{fig:internals} shows a representation of the internals of the $ACW_i$. 
The following discusses how $ACW_i$ behaves on read and write transactions. ACW is compatible with any AXI-compliant request. When $C_i$ issues a read request $AR$ through its M interface, $ACW_i$ has the following behaviors:

\begin{itemize}
\item Address Check: Check $AR$ against $LACP_i$ by comparing the address of $AR$ against each of the allowable read regions. 

\item Legal Request: If $AR$ is fully included in at least one of the allowed read regions of $LACP_i$, propagate $AR$ to the AXI interconnect.

\item Illegal Request: If $AR$ is not fully included in any of the read regions described in $LACP_i$, $AR$ is not propagated to the AXI interconnect. $ACW_i$ saves internally information regarding the illegal request $AR$. $ACW_i$ sends an AXI-compliant error to $C_i$, notifies $TE$, and switches into \emph{Decouple Mode}.

\item Outstanding Transactions: Any legal outstanding transaction initiated before an illegal transaction is completed normally. 
\end{itemize}

When $C_i$ issues a write request $AW$, $ACW_i$ behaves in a similar manner as a read request using $AW$ instead of $AR$ and comparing the address request with the write regions of $LACP_i$, instead of the read regions. 
Additionally, since AXI transactions cannot be aborted, $C_i$ expects to provide the write data after it issues the illegal request. $ACW_i$ waits to sample all the data words corresponding to $AW$ provided by $C_i$, discards them, and replies $C_i$ with an AXI-compliant error.

\subsection{IP-level Security Verification}\label{ss:IP-level-verif}

Security verification is crucial in scenarios where the design serves a security-critical role such as implementing an access control system.    
\aker uses a six-step security verification process following a property-driven hardware security methodology~\cite{hu2016towards}. The security verification process creates a threat model, identifies potential weaknesses, defines the security requirements, specifies security properties, identifies assets, and verifies the security properties. 

To drive our discussion, we first consider the IP-level verification of the ACW, which consists of three entities: a single controller $C$, a single ACW which wraps $C$, and a single peripheral $P$. We assume that the ACW’s local access control policy $LACP$ is statically configured in RTL. In later sections, we apply the same verification process to implement different SoC access control systems.  Section~\ref{sec:firmware-verification} performs  firmware-level security verification adding a hardware root-of-trust to configure the ACW. Section~\ref{sec:system-verification} describes system-level security verification that ensures that multiple ACWs  adhere to a global access control policy.

\subsubsection{Create Threat Model} The first step in the security verification process develops the threat model. It is crucial to articulate the relevant security concerns. Hardware threats are vast and must be assessed based upon the usage of the hardware under design. 
We consider an integrity scenario where $C$ is untrusted and the ACW and $P$ are trusted. Therefore, the threat model considers $C$’s ability to communicate with $P$ via the ACW in a manner which does not adhere to the statically-configured $LACP$. Threats related to confidentiality are similarly possible given that confidentiality is a dual to integrity~\cite{biba1977integrity}. 

\subsubsection{Identify Potential Weaknesses} The second step determines potential weaknesses, which is defined as any mechanism that could introduce a security vulnerability relevant to the threat model. Identifying these weaknesses is often challenging and time-consuming since it requires designers to understand the design’s specification, the design’s implementation, the subtleties in the correlation between these two, and which parts of the design are most relevant to the threat model.
In an effort to increase the chance of identifying security critical weaknesses, we use the threat model and design to find relevant CWEs from MITRE’s extensive database following to the CWE-IFT methodology~\cite{aftabjahani2021cad}.

We identified 17 CWEs relevant to the IP-level verification that are divided into two groups. The first group relates to M AXI and includes the read and write access points available to $C$, i.e., the five M AXI channels. The second group includes the configuration registers which store the ACW’s $LACP$, the anomaly registers which store illegal request metadata, and the control logic which checks the legality of and samples or blocks $C$’s transactions.

\begin{center}
\noindent\fbox{\begin{minipage}{0.95\linewidth}
\textbf{Relevant CWEs:}
1220, 1221, 1244, 1258, 1259, 1264, 1266, 1267, 1268, 1269, 1270, 1271, 1272, 1274, 1280, 1282, 1326\end{minipage}}
\end{center}

\subsubsection{Define security requirements} The third step in the process defines plain-language security requirements for the identified weaknesses. Once a mechanism is identified as a potential weakness, designers can articulate a security requirement which addresses how that mechanism could fail as determined from the relevant CWEs and an analysis of the design.

For the M AXI group of weaknesses (i.e., the AXI channels) identified in Step 2, we develop security requirements addressing the existence and the content of information flows between $C$, the ACW, and $P$. Since the ACW sits between $C$ and $P$, there will always be information flows between $C$ and the ACW and the ACW and $P$. However, the source of these flows dictates their allowable behaviors. Information flows in which the source is $C$ and the destination is $P$ (or vice versa) should only occur when the ACW is in \emph{Supervising Mode} and a legal transaction is issued. In all other instances, the only information flows that should occur are those in which the source is the ACW, the destination is either $C$ or $P$, and the content of the flow does not deviate from the default AXI values we have selected.

\begin{center}
\noindent\fbox{\begin{minipage}{0.95\linewidth}
\textbf{Requirement 1:}
C cannot receive/send data from/to P which originates while the ACW is in reset mode.\end{minipage}}
\end{center}

For the config/control group of weaknesses identified in Step 2, we develop security requirements involving the content of registers and signals. Many of the listed CWEs (e.g., 1258, 1266, 1269, and 1271)~\cite{CWE} focus on the failure to properly initialize, set, and clear the contents of security-critical registers/signals, especially on transitions between system states/modes. Considering this, the config/control group of requirements dictate what content is appropriate for registers/signals given the ACW’s current mode of operation.

\begin{center}
\noindent\fbox{\begin{minipage}{0.95\linewidth}
\textbf{Requirement 2:}
The configuration/anomaly registers are cleared and set to their default values while the ACW is actively being reset.\end{minipage}}
\end{center}

\subsubsection{Specify Security Properties} The fourth step in the process specifies a security property template for each of the security requirements. In order to verify a security requirement, it must be manually converted into a formally specified security property which uses explicit values, design signals, and operators to form an evaluable expression. Rather than specifying nearly identical security properties for each design signal that should adhere to a given security requirement, \aker provides a property generation framework which automatically generates these specific properties given a single security property template with placeholder signals and a list of target design signals.

For the security requirements relevant to the M AXI group (Requirement 1), the security property templates we specify are primarily information flow tracking $IFT$ properties. IFT properties enable us to tag information from a particular source signal and track it as it flows through our system~\cite{hu2021hardware}. For example, the send aspect of the security requirement from Requirement 1 is formalized below using the following template which fails if any information originating from $C$ during active reset flows to $P$.

\newpage

\begin{Verbatim}[frame=single]
`signal_from_C`      //source
 when (ARESETN == 0) //tagging condition
 =/=>                //no-flow operator
`signal_to_P`        //destination
\end{Verbatim}

Note that the property involves the no-flow operator (\texttt{=/=>}). Hardware information flow properties are a type of hyperproperty~\cite{clarkson2010hyperproperties}. Hyperproperties require specialized verification tools~\cite{hu2018property}. IFT properties are more challenging to verify than trace properties. Trace properties are stated over a set of traces, and are commonly used in functional verification. Hyperproperties are stated over sets of traces and are useful for proving noninterference -- a crucial aspect of information flow analysis.

For the security requirements relevant to the config/control group (Requirement 2), the security property templates are primarily trace properties which specify what the value of a specific signal/register should be under various conditions. For example, the security requirement is formalized using the following template which fails if the configuration/anomaly registers do not contain their default values after being reset.

\begin{Verbatim}[frame=single]
`reg` == `dflt_val`
 unless 
 (ARESETN != 0 && `acw_w/r_state` != 2’b00)
\end{Verbatim}

In total, we develop eighteen security property templates for verifying the security of the ACW, which are expanded to hundreds of individual properties in Step 6. Eleven of these templates are related to information flow and seven templates relate to trace properties.

\subsubsection{Identify Assets} The fifth step in the process identifies the specific assets that should adhere to each security requirement created in Step 3. Each security requirement will apply to at least one design signal (i.e., an asset) so designers must identify all of the specific assets relevant to each requirement. A single asset may be relevant for multiple security requirements.

For the security requirements related to the M AXI group, the specific assets we identify are the signals that make up the five AXI channels which connect the ACW  to $C$ and $P$. For the security requirements related to the config/control group, the specific assets we identify are the configuration and anomaly registers.

\subsubsection{Verify Security Properties} The final step generates specific security properties and then verifies them via simulation. The eighteen security property templates created in Step 4 and the assets identified in Step 5 are used to automatically generate 316 security properties which are broken down into 164 information flow properties and 152 trace properties. The verification setup for these properties includes a configurable AXI DMA module acting as controller $C$. $C$ is wrapped with an ACW. A top testbench module mimics the behavior of the peripheral $P$. The testbench iterates through resets and configurations of the ACW and the DMA with the goal of switching the ACW between operative modes to provide adequate coverage of the necessary conditions for all of the security properties.

Tortuga Logic Radix-S is used for security verification. Each property is written as an assertion using the Tortuga Logic Sentinel security language. Radix-S  generates a security model from the security rules and a simple test design. When simulated, this security model will report how many times each individual property assertion fails along with the time at which each failure occurs.

\subsection{Firmware-level Security Verification}
\label{sec:firmware-verification}

One of the key features of \aker access control systems is the simple and fast setup of the local access control policy of the ACWs. This can be setup once in static configuration (e.g., at boot time) or managed at runtime by a $TE$.  
This operation is critical. Thus, the interactions between the ACWs and the $TE$ must be securely validated.  
To perform firmware verification, we integrate \aker with the OpenTitan~\cite{OpenTitan} HRoT acting as the $TE$. The security verification proposed in this section focuses on firmware-level security verification of \aker, i.e., securely validating the interaction of the $TE$ and the ACW.
It is worth mentioning that \aker can be easily integrated with other $TE$, such as other HRoTs, trusted processors, etc.

We use the same process introduced in Section~\ref{ss:IP-level-verif} to validate the firmware-level security of the interactions between the ACW and the $TE$. There are four entities that we are concerned with: a single controller $C$, a single ACW which wraps $C$, a single peripheral $P$, and the trusted entity $TE$. Unlike the IP-level verification, the ACW’s local access control policy $LACP$ can be configured dynamically at runtime by the $TE$.

\subsubsection{Create Threat Model} In this scenario, our threat model assumes that the ACW, the $TE$, and $P$ are trusted, $C$ is untrusted, and, therefore, $C$’s ability to communicate with $P$ via the ACW in a manner which does not adhere to the dynamically-configured $LACP$ is a threat.

\subsubsection{Identify Potential Weaknesses} We identified seven relevant CWEs which helped to expand the potential weaknesses in the config/control group from Section~\ref{ss:IP-level-verif}. The additional potential weaknesses include the configuration ports which enable the $TE$ to set the ACW’s $LACP$ (i.e., the five S AXI channels) and the two interrupt lines which go from the ACW to the $TE$. Note that the CWEs and potential weaknesses identified for the IP level are still relevant for this scenario but, since we have already examined those, we only focus on weaknesses related to the interactions between the ACW and the $TE$ in this section.

\begin{center}
\noindent\fbox{\begin{minipage}{0.95\linewidth}
\textbf{Relevant CWEs:}
276, 1191, 1193, 1262, 1283, 1290, 1292\end{minipage}}
\end{center}

\subsubsection{Define security requirements} For the configuration ports identified in Step 2, we develop security requirements addressing the existence and the content of information flows between the $TE$ and the ACW’s configuration and anomaly registers. Since the $TE$ is present to configure the ACW's $LACP$, it should be the source of any information flows which modify the configuration registers. Additionally, since the anomaly registers are populated with illegal transaction metadata by the ACW for the $TE$, the $TE$ should not be able to modify the anomaly registers.

\begin{center}
\noindent\fbox{\begin{minipage}{0.95\linewidth}
\textbf{Requirement 3:}
The configuration/anomaly registers contain the default values until they are modified by the TE (config.) and/or ACW (illegal req. metadata tracking).\end{minipage}}
\end{center}

For the interrupt lines identified in Step 2, we develop security requirements addressing the value of signals. The ACW should alert the $TE$ whenever there is an illegal transaction by driving the appropriate interrupt line and otherwise it should not drive the interrupts.

\begin{center}
\noindent\fbox{\begin{minipage}{0.95\linewidth}
\textbf{Requirement 4:}
An interrupt to TE is generated after the ACW detects an illegal request.\end{minipage}}
\end{center}

\subsubsection{Specify Security Properties:} The requirements relevant to the $TE$ and the configuration/anomaly registers are primarily information flow tracking $IFT$ properties. For example, the security requirement from Requirement 3 is formalized using the following template which fails if any unauthorized source modifies the configuration regs and anomaly registers after reset.

\begin{Verbatim}[frame=single]
`unauthorized_signal`          //source
 when (`reg` == `dflt_val`)    //tagging cond.
 =/=>                          //no-flow op.
`reg`                          //destination
 unless (`reg` == `dflt_val`) 
\end{Verbatim}

For the security requirements relevant to the interrupt lines, the security property templates we specify are trace properties. In particular, Requirement 4 is formalized using the following specification which fails if the interrupt line does contain the appropriate value given the ACW's operative mode.

\begin{Verbatim}[frame=single]
`INTR_LINE_W/R` == 1
 unless (`acw_w/r_state` != 2’b10)
\end{Verbatim}

In total, we develop four security property templates for verifying the security of the firmware-level interactions between the ACW and the $TE$. Three of these are information flow properties and one is a trace property. These can all be found in our repository.

\subsubsection{Identify Assets} For the security requirements relevant to the $TE$ and the configuration/anomaly registers, the assets are all of the unauthorized signals for the configuration and anomaly registers. For the security requirements related to the ACW's interrupt lines, the assets are the read and write channel interrupt lines.

\subsubsection{Verify Security Properties} The four security property templates created in Step 4 and the assets identified in Step 5 are used to automatically generate 1,438 security properties which are broken down into 1,436 information flow properties and 2 trace properties. The verification setup for these properties is nearly identical to the IP level setup (Section~\ref{ss:IP-level-verif}) except for the presence of the $TE$ for configuring the ACW.

\subsection{System-level security verification}
\label{sec:system-verification}
Having verified the security of the ACW’s interactions at the IP level and firmware level, we now use our six-step process to validate the security of the interactions between multiple ACW-wrapped controllers and several shared SoC resources. This scenario concerns eleven entities: two controllers $C_1$ and $C_2$, two ACWs $ACW_1$ and $ACW_2$ wrapping $C_1$ and $C_2$, respectively, three peripherals $P_1 \cdots P_3$, an interconnect, and the $TE$. This scenario corresponds to an architecture from Figure~\ref{fig:extended_arch} when $N=2$ and $L=3$.
The $LACP_1$ of $ACW_1$ states that $C_1$ can read from all regions of $R_1$ = \{$P_1$, $P_2$\} and write to all regions of $W_1$ = \{$P_1$\}. The $LACP_2$ of $ACW_2$ states that $C_2$ can read from all regions of $R_2$ = \{$P_3$\} and write to all regions of $W_2$ = \{$P_2$, $P_3$\}. 

\subsubsection{Create Threat Model} In this scenario, our threat model assumes that ACWs, Ps, and the $TE$ are trusted, and that $C_1$ and $C_2$ are untrusted. The threat model focuses on the ability of the generic $C_i$’s to communicate with the generic $P_k$ via the ACW in a manner which does not adhere to the $LACP$.

\subsubsection{Identify Potential Weaknesses} We identify three additional relevant CWEs which helped to further expand the potential weaknesses from the IP and firmware verification. Since we are validating at the system level, the additional potential weaknesses include the manner in which the $LACP$ for $ACW_i$ is set as it relates the generic $C_i$ sharing resources with the generic $C_j$.   

\begin{center}
\noindent\fbox{\begin{minipage}{0.95\linewidth}
\textbf{Relevant CWEs:} 441, 1189, 1260\end{minipage}}
\end{center}

\subsubsection{Define security requirements} We develop security requirements addressing the existence and the content of information flows between every pair of $C_i$ and $P_k$ in accordance with each $ACW_i$'s $LACP_i$. $ACW_1$’s $LACP_1$ states that there should never be information flows between $C_1$ and any region of $P_3$. $ACW_2$’s $LACP$ states that there should never be information flows between $C_2$ and any region of $P_1$. 

\begin{center}
\noindent\fbox{\begin{minipage}{0.95\linewidth}
\textbf{Requirement 5:}
Any C cannot receive/send data from/to any region not contained within its ACW’s LACP.\end{minipage}}
\end{center}

\subsubsection{Specify Security Properties:}The security property templates we specify are all $IFT$ properties. For example, the send aspect of Requirement 5  is formalized below using the following template which fails if any information originating from some $C_i$ flows to any unauthorized region.

\begin{Verbatim}[frame=single]
`sig_from_C`   //source (always tagged)
 =/=>
`unauthorized` //destination
\end{Verbatim}

In total, we develop two $IFT$ security property templates for verifying the security of the interactions between multiple ACW-wrapped controllers and multiple shared resources across our system.

\subsubsection{Identify Assets}The specific assets we identify as being relevant for our security requirements are all of the signals within the unauthorized regions for each $C_i$ and the AXI signals which connect each $C_i$ to its respective $ACW_i$ 

\subsubsection{Verify Security Properties} The two security property templates created in step 4 and the assets identified in step 5 are used to automatically generate 76 $IFT$ security properties. The verification setup for these properties builds upon the Firmware-level verification from Section~\ref{sec:firmware-verification} by inserting one additional ACW-wrapped controller module, three memory modules to serve as peripherals, and one AXI interconnect to connect the controllers with the peripherals.

\section{Experimental evaluation}\label{s:exper}
We evaluate the performance and resource usage of the \aker access control system and verify its functional and security correctness.
The first set of experiments in Section~\ref{s:fpga-exper} compares the performance and resource usage of an \aker access control system with two typical methods of implementing on-chip access control described in Section~\ref{ss:SoA-solutions}.
In Section~\ref{s:PULP}, we provide a case study of integrating an \aker access control system on the OpenPULP architecture~\cite{conti2016pulp}.
\subsection{FPGA SoC Experiments}\label{s:fpga-exper}
We develop an FPGA SoC architecture on a Xilinx Zynq Ultrascale+ platform. The architecture has three controller modules $C_1$, $C_2$, and $C_3$ implemented as hardware accelerators in the FPGA fabric. $C_1$, $C_2$, and $C_3$ are connected to a Xilinx AXI SmartConnect~\cite{smartconnect}, which is connected to a single peripheral resource $P_1$ -- the shared DRAM memory controller exposed as an AXI S bus in the Processing System $PS$. This architecture is similar to the one shown in  Figure~\ref{fig:generic_arch} with $N=3$ and $L=1$.
$C_1$, $C_2$, and $C_3$ are implemented as three separate high-performance DMA IPs -- this choice allows easily configuration and can cover a wide range of communication behaviors.

We implemented three different access control system: \textbf{Design (a)} uses AXI SmartConnect (INTC) (Section~\ref{sss:AXI-intc}), \textbf{Design (b)} implements access control in $PS$ leveraging the Xilinx Memory Protection Unit (XMPU)~\cite{XMPU} (Section~\ref{sss:peripheral-res}), and \textbf{Design (c)} is an \aker access control system involving three ACW modules. We leverage one of the processors of the platform for the configuration of the ACWs, which acts as the $TE$. 
The designs have been synthesized using Xilinx Vivado 2018.2.
To achieve high accuracy, the performance measurements are done by a custom timer implemented into the FPGA fabric. We also deployed a Xilinx System ILA~\cite{xil-ila} to verify the correct behavior of the ACW modules.

\begin{table}[]
\footnotesize
\begin{tabular}{|l|c|c|c|}
\hline
\textbf{Functionality for policy definition}                                                & \textbf{INTC} & \textbf{XMPU} & \textbf{\aker} \\ \hline
Protect a limited set of predefined regions                              & Yes  & Yes  & \textbf{Yes} \\ \hline
Dynamic allocation of read/write regions                     & No   & Yes  & \textbf{Yes} \\ \hline
Definition of private read/write regions                            & No   & No  & \textbf{Yes} \\ \hline
Definition of read-only/write-only regions                  & No   & No   & \textbf{Yes} \\ \hline
Secure transactions identification                  & Yes   & No   & \textbf{Yes} \\ \hline
\end{tabular}
\caption{Comparison of SoC access control systems. The AXI-interconnect-based INTC and the XMPU access control systems available exhibit limitations not seen in with \aker.}
\label{tab:policies-features}
\end{table}

One of the major drawbacks of the access control system implemented with the AXI SmartConnect in Design \textbf{(a)} is that the access control policy cannot be dynamically modified. Moreover, the Vivado design tool uses predefined addressable regions for the controllers -- once set, no additional custom regions can be added in the access control policy of the AXI SmartConnect. 
Also, the predefined regions have default read/write permissions -- no read only regions can be defined. 

Design \textbf{(b)} uses the XMPU integrated into the $PS$ to implement the access control system, which enables the definition of up to 16 custom memory regions.
The XMPU uses the workaround of the AXI ID signals described in Section~\ref{sss:peripheral-res} to identify the source of a specific request and decide whether a request is legal or not (see~\cite{XMPU}).
However, the AXI SmartConnect connecting the $C$ to the PS does not propagate the ID signals to the PS~\cite{smartconnect}. 
Therefore, even when forcing $C_1$, $C_2$, and $C_3$ to issue requests with unique IDs, once the requests are propagated by the AXI SmartConnect to the PS, they lose the ID and therefore the XMPU cannot determine the source of the request. 
Thus, the XMPU cannot reliably enforce any access control policy that aims to differentiate the requests issued by $C_1$, $C_2$, and $C_3$.
This means that even a very simple access control policy defining a private read/write buffer for each $C$ cannot be implemented using the XMPU as the access control system.
Also in this case, the regions defined in the XMPU has default read/write privilege that cannot be target for the definition of read-only regions.
Table~\ref{tab:policies-features} summarizes some of the features required to the access control system for the implementation of common functionalities required in security policies and the limitations of the methods available on the modern SoC under evaluation.
We evaluate the \aker access control system with Designs \textbf{(a)} and \textbf{(b)} to understand relative performance and resource usage. Therefore, we develop a simple access control policy compatible with the limitations of the access control systems of designs \textbf{(a)} and \textbf{(b)} and compare these three designs.

\begin{figure}[htb!]
\centering
\begin{tikzpicture}
\tikzstyle{every node}=[font=\footnotesize]
\begin{groupplot}[
    group style = {group size=3 by 3, horizontal sep=1cm, vertical sep=1.2cm},
    width=0.4\columnwidth,
    height=3.5cm,
    legend style={at={(-1,-0.4)}, anchor=north,legend columns=-1},
]
\nextgroupplot[
    title  = {},
    title style={yshift=-6pt},
    ylabel = {Resp. times ($\mu s$)},
    ybar=2pt,
    bar width=5pt,
    	xtick={0, 1},
	xticklabels={16-word, 256-word},
    xtick=data,
        axis lines* = left,
    ymax=9,
        enlarge x limits  = 0.4,
    ymajorgrids=true,
    ]
\addplot[black, fill=cb1]
	coordinates {
		(0, 1.22)
		(1, 5.92)
	};

\addplot[black, fill=cb2]
	coordinates {
		(0, 1.21)
		(1, 5.93)
	};

\addplot[black, fill=cb3]
	coordinates {
		(0, 1.23)
		(1, 5.95)
	};

\nextgroupplot[
    title  = {\textbf{(i)} Performance of the DMAs in isolation},
    title style={yshift=-6pt},
    ybar=2pt,
    bar width=5pt,
    	xtick={0, 1},
	xticklabels={4 KB, 32 KB},
    xtick=data,
        axis lines* = left,
    ymax=110,
        enlarge x limits  = 0.4,
    ymajorgrids=true,
    ]
\addplot[black, fill=cb1]
	coordinates {
		(0, 10.86)
		(1, 82.41)
	};

\addplot[black, fill=cb2]
	coordinates {
		(0, 10.85)
		(1, 82.43)
	};

\addplot[black, fill=cb3]
	coordinates {
		(0, 10.89)
		(1, 82.85)
	};

\nextgroupplot[
    title  = {},
    title style={yshift=-6pt},
    ybar=2pt,
    bar width=5pt,
    	xtick={0, 1},
	xticklabels={256 KB, 2 MB},
    xtick=data,
        axis lines* = left,
    ymax=7000,
        enlarge x limits  = 0.4,
    ymajorgrids=true,
    ]
\addplot[black, fill=cb1]
	coordinates {
		(0, 655.92)
		(1, 5226.31)
	};

\addplot[black, fill=cb2]
	coordinates {
		(0, 656.13)
		(1, 5227.10)
	};

\addplot[black, fill=cb3]
	coordinates {
		(0, 658.92)
		(1, 5243.34)
	};

\nextgroupplot[
  title  = {},
  title style={yshift=-6pt},
  ylabel = {Resp. times ($\mu s$)},
  ybar=2pt,
  bar width=5pt,
	xtick={0, 1},
	xticklabels={16-word, 256-word},
  xtick=data,
    axis lines* = left,
  ymax=9,
    enlarge x limits  = 0.4,
  ymajorgrids=true,
  ]
\addplot[black, fill=cb1]
	coordinates {
		(0, 1.22)
		(1, 5.92)
	};

\addplot[black, fill=cb2]
	coordinates {
		(0, 3.7)
		(1, 8.3)
	};

\addplot[black, fill=cb3]
	coordinates {
		(0, 1.22)
		(1, 5.95)
	};
	
\nextgroupplot[
  title  = {\textbf{(ii)} $C_2$ interfere the execution of $C_1$ issuing illegal transactions},
  title style={yshift=-6pt},
  ybar=2pt,
  bar width=5pt,
	xtick={0, 1},
	xticklabels={4 KB, 32 KB}, 
  xtick=data,
    axis lines* = left,
    enlarge x limits  = 0.4,
  ymajorgrids=true,
  ]
\addplot[black, fill=cb1]
	coordinates {
		(0, 10.86)
		(1, 82.55)
	};

\addplot[black, fill=cb2]
	coordinates {
		(0, 12.96)
		(1, 100.61)
	};

\addplot[black, fill=cb3]
	coordinates {
		(0, 10.89)
		(1, 82.91)
	};
	
\nextgroupplot[
  title  = {},
  title style={yshift=-6pt},
  ybar=2pt,
  bar width=5pt,
	xtick={0, 1},
	xticklabels={256 KB, 2 MB},
  xtick=data,
    axis lines* = left,
    enlarge x limits  = 0.4,
  ymajorgrids=true,
  ]
\addplot[black, fill=cb1]
	coordinates {
		(0, 655.92)
		(1, 5226.31)
	};

\addplot[black, fill=cb2]
	coordinates {
		(0, 816.80)
		(1, 6551.16)
	};

\addplot[black, fill=cb3]
	coordinates {
		(0, 658.92)
		(1, 5243.34)
	};
	
\nextgroupplot[
  title  = {},
  title style={yshift=-6pt},
  ylabel = {Resp. times ($\mu s$)},
  ybar=2pt,
  bar width=5pt,
	xtick={0, 1},
	xticklabels={16-word, 256-word},
  xtick=data,
    axis lines* = left,
    enlarge x limits  = 0.4,
  ymajorgrids=true,
  ]
\addplot[black, fill=cb1]
	coordinates {
		(0, 1.22)
		(1, 5.92)
	};

\addplot[black, fill=cb2]
	coordinates {
		(0, 38.72)
		(1, 40.96)
	};

\addplot[black, fill=cb3]
	coordinates {
		(0, 1.22)
		(1, 5.92)
	};

\nextgroupplot[
  title  = {\textbf{(iii)} $C_2$ acts a DoS on $C_1$ flooding the bus with illegal transactions},
  title style={yshift=0pt},
  ybar=2pt,
  bar width=5pt,
	xtick={0, 1},
	xticklabels={4 KB, 32 KB},
  xtick=data,
    axis lines* = left,
    enlarge x limits  = 0.4,
  ymajorgrids=true,
  ]
\addplot[black, fill=cb1]
	coordinates {
		(0, 10.78)
		(1, 82.95)
	};

\addplot[black, fill=cb2]
	coordinates {
		(0, 48.71)
		(1, 222.27)
	};

\addplot[black, fill=cb3]
	coordinates {
		(0, 10.95)
		(1, 84.01)
	};
	
\nextgroupplot[
  title  = {},
  title style={yshift=-6pt},
  ybar=2pt,
  bar width=5pt,
	xtick={0, 1},
	xticklabels={256 KB, 2 MB},
  xtick=data,
    axis lines* = left,
    enlarge x limits  = 0.4,
  ymajorgrids=true,
  ]
\addplot[black, fill=cb1]
	coordinates {
		(0, 655.92)
		(1, 5226.31)
	};

\addplot[black, fill=cb2]
	coordinates {
		(0, 1742.89)
		(1, 13910.92)
	};

\addplot[black, fill=cb3]
	coordinates {
		(0, 658.92)
		(1, 5243.34)
	};

\legend{(a) AXI INTC \quad\quad, (b) XMPU \quad\quad, (c) \aker}
\end{groupplot}
\end{tikzpicture}
\caption{Performance Evaluation: \textbf{(i)}: the three designs act similarly in isolation. \textbf{(ii)}: The techniques differ in response time in situations with illegal transaction requests.  \textbf{(iii)}: A DOS attack by $C_2$ endangers the availability of the DRAM memory from $C_1$ in Design \textbf{(b)}.}
\label{fig:perf}
\end{figure}
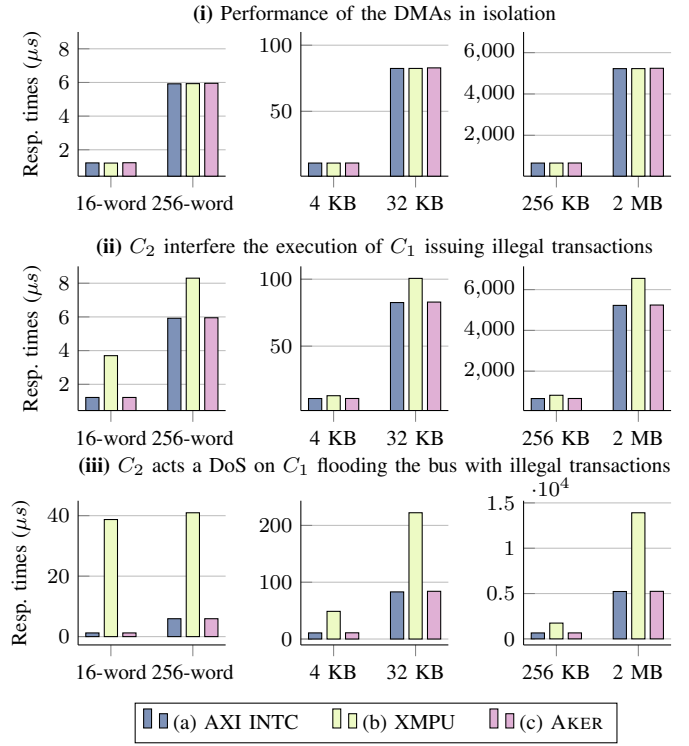

The first experiment compares the performance impact on the memory access time of $C_1$, $C_2$, and $C_3$ associated with the three access control systems. 
We setup a common forbidden region $F$ in memory -- any read or write request directed to that region is illegal for any controllers $C$.
We first evaluate the memory access time in isolation: $C_1$, $C_2$, and $C_3$ are activated separately and access a legal region of the memory for different amounts of data. Figure~\ref{fig:perf}(i) reports the measured memory access times.
The results show similar performance in latency and throughput for all the designs. This result confirms that the per-transaction extra one clock cycle introduced by the ACW modules has a negligible impact on performance. 

The next experiment tests contention generated by illegal transactions. We keep the same configuration of the previous experiment for $C_1$, but we configure $C_2$ to try to concurrently access the forbidden region $F$, i.e., it issues illegal requests.  $C_2$ issues a new illegal transaction once the previous one has been replied with an error -- this behavior models a controller stuck trying to access an illegal memory region due to a misconfiguration. 
As discussed in Section~\ref{sss:peripheral-res}, illegal requests in access control systems implemented at the peripheral (Design \textbf{(b)}) occupy time on the interconnect and therefore impact the performance of legal transactions.
The measured average access time for $C_1$ are reported in Figure~\ref{fig:perf}(ii).
The results from Design \textbf{(b)} show that the interference generated by the illegal transactions issued by $C_2$ impacts the performance of $C_1$. In particular, the average measured response time increases by 203\% on a 16-word transaction changing from 1.22 $\mu \, s$ in Designs \textbf{(a)} and \textbf{(c)} to 3.7 $\mu \,s$ in Design \textbf{(c)}. 
We measured a lower impact for longer and consecutive accesses. However, in all the cases we measured an impact of at least 20\% on the average response times.
In Design \textbf{(a)} and \textbf{(c)} the illegal transactions of $C_2$ are stopped before entering the network -- the results shows how \aker stops any interference generated by illegal transactions while featuring flexibility in the definition of the access control policy. 

The final experiment tests a denial of service scenario. $C_1$ keeps the same configuration from the previous experiment while $C_2$ is setup to flood the interconnect, issuing continuous illegal requests, thus leveraging the full throughput made available by the AXI SmartConnect to $C_2$. In this case, $C_2$ mimics the behaviour of a misconfigured or malicious high-throughput IP core.
The results are reported in Figure~\ref{fig:perf}(iii). The impact on the response times of $C_1$ in Design \textbf{(b)} is way higher than in part \textbf{(ii)}: the average measured response time of a 16-word transaction issued by $C_1$ passes from 1.22 $\mu \, s$ of Designs \textbf{(a)} and \textbf{(c)} to 38.72 $\mu \, s$ in Design \textbf{(b)}, corresponding to an increase of 3074\%.
Again, the impact decreases on longer and consecutive accesses, however, all of the tested cases showed an impact of at least the 165\% on the nominal average response times. Thus, in all the tested scenarios, the response time of $C_1$ is more than the double with respect to nominal conditions.
This experiment shows how a misbehaving IP can create a denial of service when using Design \textbf((b) for access control. 
This issue can be critical in designs integrating software-configurable IPs -- malicious software could exploit this issue to act Denial-of-Service of the memory or other resources to the other IPs integrated into the system.
Indeed, even if detected at runtime, the access control system implemented in Design \textbf{(b)} does not provide any method to stop the flood of illegal transactions.

\begin{table}[h]
\centering
\footnotesize
\begin{tabular}{|c|c|c|c|c|c|}
\hline
Resources & \textbf{AXI SMTC} & \textbf{2 regs} & \textbf{4 regs} & \textbf{8 regs} & \textbf{16 regs} \\ \hline
\textbf{LUT}    &  5626 &  263 & 326   &  467  & 730 \\ \hline
\textbf{FF}     &  6688 &  294 &  358  &  486  & 744 \\ \hline
\end{tabular}
\caption{Resource consumption of the ACW module. The area impact of the ACW can be target according to the requirements of the target application.}
\label{tab:footprint}
\end{table}

Table~\ref{tab:footprint} reports the resource consumption for different configurations of the ACW module implemented for the Xilinx ZYNQ Ultrascale+ platform. The results report the resource consumption of ACW deploying 2 regions (2 regs), 4 regions (4 regs), 8 regions (8 regs), and 16 regions (16 regs). The results are compared with the resource cost of the AXI SmartConnect used in Design \textbf(c) (AXI SMTC). 
The results show that the ACW modules have a limited impact on resource consumption with respect to the cost of the AXI SmartConnect. Moreover, the resource consumption can optimized to met the requirements of the application.

\subsection{PULP SoC Experiments}\label{s:PULP}

The Parallel Ultra-Low-Power (PULP) is an open-source multi-core computing platform comprised of a multicore RISC-V processor. PULP is divided into the SoC domain and the Cluster domain where the SoC performs control and other high level functions while the cluster is aimed at hardware acceleration incorporating eight RISC-V cores. 

There are two communication pathways between the SoC and the Cluster. One pathway allows the Cluster to access the SoC, i.e., the Cluster is the Controller and the SoC is a Peripheral. 
The other communication pathway allows the SoC  to access the Cluster -- the SoC is the Controller and the Cluster is a Peripheral. 
These two communication pathways enable the fabric controller core in the SoC domain and the eight cores in the Cluster domain to send/receive information and access the shared L2 memory. Additionally, the PULP’s memory map which includes areas for the Cluster subsystem, the ROM memory, the SoC peripherals subsystem, and the L2 memory is fully addressable from any of the PULP’s nine cores. 

The OpenPULP platform does not feature any default access control system for the transactions going from the cluster to the SoC, and vice-versa.
Thus, we integrated an \aker access control system using two ACWs to regulate the communication between the PULP’s SoC domain and Cluster domain. Since both pathways use the AXI standard, the process of wrapping their respective AXI M is straightforward and only requires ensuring that the ACW’s ports are connected to the proper signals. 
Once this is completed, the two ACWs filter all the read/write transactions on the pathways. 

\begin{figure}[htb!]
\centering
\includegraphics[width=0.99\columnwidth]{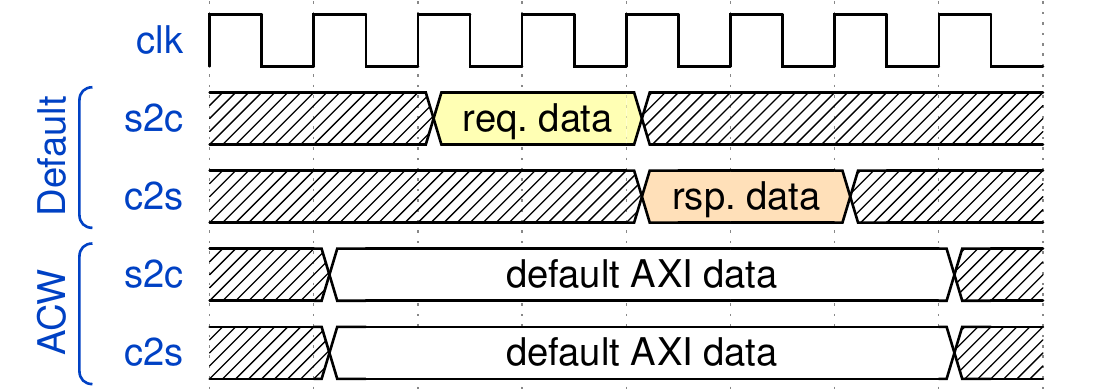}
\caption{\textcolor{blue}{s2c} is the SoC to Cluster pathway. \textcolor{blue}{c2s} is the Cluster to SoC pathway. The \textcolor{blue}{Default} group shows a portion of the execution of a C program using the Cluster with the default OpenPulp. The \textcolor{blue}{ACW} group shows the same portion of the execution with the addition of two ACWs configured to block all transactions.}
\label{fig:OpenPulp}
\end{figure}

We validated the access control policies enforced by the two ACWs via test simulations with various ACW configurations. The test simulations we used include C programs that do not make use of the PULP Cluster and C programs which do make use of the PULP Cluster. 
As a baseline, we ensured that all test simulations are able to successfully run on the default PULP. 
For our first validation, we configured both ACWs to allow all read/write transactions and verify that all tests run successfully as with the baseline. For our second validation, we configured both ACWs to block all read/write transactions and verify that the tests which do not make use of the PULP Cluster run successfully and those that do use the Cluster stall while waiting for responses from the decoupled domains. 
For the remaining validations, we used a combination of  configurations to ensure the ACWs are able to enforce access control policies that are more fine-grained than the all-or-nothing approach used in the first two validations. 
For each validation, we verified results using the testbench output logs and vcd/waveforms (see Figure~\ref{fig:OpenPulp}).  %
The design and testing frameworks are available in our repository \cite{Github}.

\section{Related Works}\label{ss:related-works}

Access control systems have been integrated into network-on-chip (NoC) architectures.  Fiorin et al. study different manners of integration between centralized and distributed~\cite{fiorin2008secure}. Grammatikakis et al. describe a NoC firewall that checks memory accesses in a distributed manner~\cite{grammatikakis2015security}. SurfNoC~\cite{wassel2013surfnoc} can isolate mixed-trust users and effectively utilizing the NoC in a time-multiplexed manner. Sepulveda et al.~\cite{Sepulveda2018} proposed a property-driven method aimed at the specific security verification of NoC routers. 
While \aker did not specifically focus on NoC architectures, it could be extended to them with some modifications to our methodology, opening the possibility for interesting future works. 

Several research efforts have been spent by the research community to advance the security of shared bus architectures.
Jacob et al.~\cite{Jacob2017} demonstrated on a commercial platform how hardware vulnerabilities can be injected in real systems during the integration of third-party IP modules (3PIP), also relating to access control. A brief discussion of prevention techniques are provided without providing any specific solution. 
Siddiqui et al.~\cite{siddiqui2018pro} and Tan et al.~\cite{tan2020towards} proposed two solutions for the implementation of distributed and decentralized systems aiming to detect anomalous conditions generated by hardware modules. While these solutions can help mitigating misbehaving conditions generated by the hardware module, they are not intended for the implementation of dynamic access control systems.
Brunel et al.~\cite{Brunel2014} provided a software/hardware system for securing the off-chip memories with static policies during boot phase. It cannot handle dynamic access control policies.
Cotret et al.~\cite{Cotret2012} proposed an hardware module for the deployment of distributed firewall systems. 
This has similarities with our proposed method. However, it lacks a security verification strategy, it shows high performances impact (18\% increase in latency, that can also depends on how the region table is sorted), and lacks of any integration with modern HRoTs for secure configuration.

\section{Conclusions}
We developed \aker~-- a design and verification framework for SoC access control systems. \aker builds upon an efficient and modular Access Control Wrapper (ACW) that easily integrates with on-chip controller resources to monitor their memory transactions. \aker provides an extensive security verification framework that follows a property-driven verification methodology. We show how to design, integrate, and verify \aker access control systems into several SoC architectures. We demonstrate that \aker has limited impact on performance while using minimal resources. \aker is easily integrated with a hardware root-of-trust to ensure secure configuration of the the local ACW access control policies. \aker is released as open-source repository~\cite{Github}. This includes and extensive verification framework with security templates, properties, and testbenches to perform security verification at the IP, firmware, and system levels. The repository also includes design files for the ACW and example integration with OpenTitan, OpenPULP, and a Xilinx FPGA SoC.

\begin{spacing}{0.5}
\bibliographystyle{IEEEtran}
\bibliography{biblio.bib}

\begin{thebibliography}{10}
\providecommand{\url}[1]{#1}
\csname url@samestyle\endcsname
\providecommand{\newblock}{\relax}
\providecommand{\bibinfo}[2]{#2}
\providecommand{\BIBentrySTDinterwordspacing}{\spaceskip=0pt\relax}
\providecommand{\BIBentryALTinterwordstretchfactor}{4}
\providecommand{\BIBentryALTinterwordspacing}{\spaceskip=\fontdimen2\font plus
\BIBentryALTinterwordstretchfactor\fontdimen3\font minus
  \fontdimen4\font\relax}
\providecommand{\BIBforeignlanguage}[2]{{%
\expandafter\ifx\csname l@#1\endcsname\relax
\typeout{** WARNING: IEEEtran.bst: No hyphenation pattern has been}%
\typeout{** loaded for the language `#1'. Using the pattern for}%
\typeout{** the default language instead.}%
\else
\language=\csname l@#1\endcsname
\fi
#2}}
\providecommand{\BIBdecl}{\relax}
\BIBdecl

\bibitem{CWE}
\emph{The Common Weakness Enumeration Official Webpage}, MITRE,
  https://cwe.mitre.org/.

\bibitem{hu2016towards}
W.~Hu, A.~Althoff, A.~Ardeshiricham, and R.~Kastner, ``Towards property driven
  hardware security,'' in \emph{2016 17th International Workshop on
  Microprocessor and SOC Test and Verification (MTV)}.\hskip 1em plus 0.5em
  minus 0.4em\relax IEEE, 2016, pp. 51--56.

\bibitem{OpenTitan}
\emph{The OpenTitan Hardware Root of Trust offical website}, OpenTitan,
  https://opentitan.org/.

\bibitem{Github}
\emph{Aker Anonymous Github}, https://github.com/dsfkl/Aker.git.

\bibitem{conti2016pulp}
F.~Conti, D.~Rossi, A.~Pullini, I.~Loi, and L.~Benini, ``Pulp: A ultra-low
  power parallel accelerator for energy-efficient and flexible embedded
  vision,'' \emph{Journal of Signal Processing Systems}, vol.~84, no.~3, pp.
  339--354, 2016.

\bibitem{smartconnect}
\emph{AXI SmartConnect v1.0 LogiCORE IP Product Guide}, Xilinx, pG247.

\bibitem{siddiqui2018pro}
F.~Siddiqui, M.~Hagan, and S.~Sezer, ``Pro-active policing and policy
  enforcement architecture for securing mpsocs,'' in \emph{2018 31st IEEE
  International System-on-Chip Conference (SOCC)}.\hskip 1em plus 0.5em minus
  0.4em\relax IEEE, 2018, pp. 140--145.

\bibitem{coburn2005seca}
J.~Coburn, S.~Ravi, A.~Raghunathan, and S.~Chakradhar, ``Seca:
  security-enhanced communication architecture,'' in \emph{Proceedings of the
  2005 international conference on Compilers, architectures and synthesis for
  embedded systems}, 2005, pp. 78--89.

\bibitem{XMPU}
\emph{Isolation Methods in Zynq UltraScale+ MPSoCs}, Xilinx, xAPP1320.

\bibitem{amba4spec}
\emph{AMBA® AXI™ and ACE™ Protocol Specification}, ARM, iHI 0022D.

\bibitem{basak2015flexible}
A.~Basak, S.~Bhunia, and S.~Ray, ``A flexible architecture for systematic
  implementation of soc security policies,'' in \emph{2015 IEEE/ACM
  International Conference on Computer-Aided Design (ICCAD)}.\hskip 1em plus
  0.5em minus 0.4em\relax IEEE, 2015, pp. 536--543.

\bibitem{nath2018system}
A.~P.~D. Nath, S.~Ray, A.~Basak, and S.~Bhunia, ``System-on-chip security
  architecture and cad framework for hardware patch,'' in \emph{2018 23rd Asia
  and South Pacific Design Automation Conference (ASP-DAC)}.\hskip 1em plus
  0.5em minus 0.4em\relax IEEE, 2018, pp. 733--738.

\bibitem{biba1977integrity}
K.~J. Biba, ``Integrity considerations for secure computer systems,'' MITRE
  CORP BEDFORD MA, Tech. Rep., 1977.

\bibitem{aftabjahani2021cad}
S.~Aftabjahani, R.~Kastner, M.~Tehranipoor, F.~Farahmandi, J.~Oberg,
  A.~Nordstrom, N.~Fern, and A.~Althoff, ``Cad for hardware security -
  automation is key to adoption of solutions,'' in \emph{Proceedings of the
  IEEE VLSI Test Symposium}, 2021.

\bibitem{hu2021hardware}
W.~Hu, A.~Ardeshiricham, and R.~Kastner, ``Hardware information flow
  tracking,'' \emph{ACM Computing Surveys (CSUR)}, vol.~54, no.~4, pp. 1--39,
  2021.

\bibitem{clarkson2010hyperproperties}
M.~R. Clarkson and F.~B. Schneider, ``Hyperproperties,'' \emph{Journal of
  Computer Security}, vol.~18, no.~6, pp. 1157--1210, 2010.

\bibitem{hu2018property}
W.~Hu, A.~Ardeshiricham, M.~S. Gobulukoglu, X.~Wang, and R.~Kastner, ``Property
  specific information flow analysis for hardware security verification,'' in
  \emph{Proceedings of the International Conference on Computer-Aided Design},
  2018, pp. 1--8.

\bibitem{xil-ila}
\emph{System Integrated Logic Analyzer v1.0}, Xilinx, 2017, pG261.

\bibitem{fiorin2008secure}
L.~Fiorin, G.~Palermo, S.~Lukovic, V.~Catalano, and C.~Silvano, ``Secure memory
  accesses on networks-on-chip,'' \emph{IEEE Transactions on Computers},
  vol.~57, no.~9, pp. 1216--1229, 2008.

\bibitem{grammatikakis2015security}
M.~D. Grammatikakis, K.~Papadimitriou, P.~Petrakis, A.~Papagrigoriou,
  G.~Kornaros, I.~Christoforakis, O.~Tomoutzoglou, G.~Tsamis, and M.~Coppola,
  ``Security in mpsocs: a noc firewall and an evaluation framework,''
  \emph{IEEE Transactions on Computer-Aided Design of Integrated Circuits and
  Systems}, vol.~34, no.~8, pp. 1344--1357, 2015.

\bibitem{wassel2013surfnoc}
H.~M. Wassel, Y.~Gao, J.~K. Oberg, T.~Huffmire, R.~Kastner, F.~T. Chong, and
  T.~Sherwood, ``Surfnoc: a low latency and provably non-interfering approach
  to secure networks-on-chip,'' \emph{ACM SIGARCH Computer Architecture News},
  vol.~41, no.~3, pp. 583--594, 2013.

\bibitem{Sepulveda2018}
J.~Sepulveda, D.~Aboul-Hassan, G.~Sigl, B.~Becker, and M.~Sauer, ``Towards the
  formal verification of security properties of a network-on-chip router,'' in
  \emph{2018 IEEE 23rd European Test Symposium (ETS)}.\hskip 1em plus 0.5em
  minus 0.4em\relax IEEE, 2018, pp. 1--6.

\bibitem{Jacob2017}
N.~{Jacob}, C.~{Rolfes}, A.~{Zankl}, J.~{Heyszl}, and G.~{Sigl}, ``Compromising
  fpga socs using malicious hardware blocks,'' in \emph{Design, Automation Test
  in Europe Conference Exhibition (DATE), 2017}, 2017, pp. 1122--1127.

\bibitem{tan2020towards}
B.~Tan, R.~Elnaggar, J.~M. Fung, R.~Karri, and K.~Chakrabarty, ``Towards
  hardware-based ip vulnerability detection and post-deployment patching in
  systems-on-chip,'' \emph{IEEE Transactions on Computer-Aided Design of
  Integrated Circuits and Systems}, 2020.

\bibitem{Brunel2014}
J.~{Brunel}, R.~{Pacalet}, S.~{Ouaarab}, and G.~{Duc}, ``Secbus, a
  software/hardware architecture for securing external memories,'' in
  \emph{2014 2nd IEEE International Conference on Mobile Cloud Computing,
  Services, and Engineering}, 2014, pp. 277--282.

\bibitem{Cotret2012}
P.~{Cotret}, J.~{Crenne}, G.~{Gogniat}, and J.~{Diguet}, ``Bus-based mpsoc
  security through communication protection: A latency-efficient alternative,''
  in \emph{2012 IEEE 20th International Symposium on Field-Programmable Custom
  Computing Machines}, 2012, pp. 200--207.

\end{thebibliography}
\end{spacing}
\end{document}